\documentclass[showpacs,preprintnumbers]{revtex4}
\usepackage{amsmath}
\usepackage{txfonts}
\usepackage{amssymb}
\usepackage[mathscr]{eucal}
\usepackage[dvips]{color}

\input{tcilatex}
\begin{document}

\title{Anomalies of the Achucarro-Ortiz black hole}
\author{Kui Xiao}
\email{87xiaokui@163.com}
\author{Wenbiao Liu}
\email{wbliu@bnu.edu.cn}
\affiliation{Department of Physics, Institute of Theoretical Physics, Beijing Normal
University, Beijing, 100875, China}
\author{Hongbao Zhang}
\email{hbzhang@pkuaa.edu.cn}
\affiliation{Department of Astronomy, Department of Physics, Beijing Normal University,
Beijing, 100875, China\\
CCAST (World Laboratory), P.O. Box 8730, Beijing, 100080, China}

\begin{abstract}
Considering anomalies of quantum field in the $(1+1)-$dimensional
Achucarro-Ortiz black hole background, the stress tensor near and out of the
horizon is calculated, meanwhile, the relationship between anomalies and
Hawking radiation of the black hole is discussed.

Keywords: Achucarro-Ortiz black hole; stress tensor; confromal anomaly;
gauge anomaly; gravitation anomaly; Hawking radiation
\end{abstract}

\pacs{04.62.+v, 04.70.Dy, 11.30.--j }
\maketitle

\section{Introduction}

Anomalies are the breaking of classical symmetries by quantum mechanics, and
these phenomena are always used to understand a new physics system \cite%
{harvey}. Conformal anomaly appears in a theory that the energy momentum has
a trace $T_{\alpha }^{\alpha }$ which vanishes classically. The relation
between conformal anomaly and Hawking radiation \cite{Hawking1} has been
studied by many authors \cite{Christensen,Lohiya,Frusaev}. Considering
conformal anomaly, many scientists studied the Casimir effect in
(1+1)-dimensional curved spacetime background \cite{setare,gao}.

Gravitational anomaly shows the nonconservation of stress tensor. In
(1+1)-dimensional spacetime, the gravitational anomaly is given by \cite%
{Bertlmann} 
\begin{equation}
\nabla _{\mu }T_{\nu }^{\mu }=\frac{1}{96\pi \sqrt{-g}}\epsilon ^{\beta
\delta }\partial _{\delta }\partial _{\alpha }\Gamma _{\nu \beta }^{\alpha }=%
\frac{1}{\sqrt{-g}}\partial _{\mu }N_{\nu }^{\mu }=\mathcal{A}_{\nu }.
\label{eq1}
\end{equation}%
In this field, Robinson and Wilczek discussed the gravitational anomaly in
the region that localized on the width of $2\varepsilon $ straddling the
horizon and found that the flux of Hawking radiation cancels the
gravitational anomaly \cite{Robinson}. Based on the work of \cite{Robinson},
Iso et al calculated the relation between the Hawking radiation of a charged
black hole and the gravitational anomaly and gauge anomaly \cite{Iso}. Many
other authors discussed the anomalies near the horizon of different black
holes, ie, rotating black hole \cite{Iso2,Keiju}, dynamical black hole \cite%
{Vagenas1} and $(2+1)$-dimensional black hole \cite{setare2}.

In this paper, we will calculate the anomalies in $(1+1)$-dimensional
Achucarro-Ortiz black hole background. It is organized as following. In
Section II, we will introduce the Achucarro-Ortiz black hole in detail. In
Section III, the stress tensor in the region ($r_{+}+\varepsilon <r<\infty $%
) is calculated, and then the contribution of trace anomaly to the stress
tensor is discussed. In Section IV, we study the gravitational anomaly and
gauge anomaly near the horizon, meanwhile, the relation between Hawking
radiation and anomalies in this region is discussed. In Section V, we will
get conclusions and give some comments.

\section{Achucarro-Ortiz black hole}

The (2+1)-dimensional BTZ black hole was first derived by Banados
Teitelboim, and Zanelli \cite{Banados} under the action 
\begin{equation}
S=\int d^{3}x\sqrt{-g}(R+2\Lambda ),
\end{equation}%
with a negative cosmology constant $\Lambda $. The element of BTZ black hole
is 
\begin{eqnarray}
ds^{2} &=&-(-M+\Lambda r^{2}+\frac{J^{2}}{4r^{2}})dt^{2}+(-M+\Lambda r^{2}+%
\frac{J^{2}}{4r^{2}})^{-1}dr^{2}  \notag \\
&&+r^{2}(d\theta -\frac{J}{2r^{2}}dt)^{2}.
\end{eqnarray}

There are many ways to reduce this line element to (1+1)-dimensional one 
\cite{Achucarro-Ortiz,Kumar}. Under the Kaluza-Klein reduction the metric of
BTZ black hole yields the two-dimensional line element as \cite{Vagenas} 
\begin{equation}
ds^{2}=-(-M+\Lambda r^{2}+\frac{J^{2}}{4r^{2}})dt^{2}+(-M+\Lambda r^{2}+%
\frac{J^{2}}{4r^{2}})^{-1}dr^{2},  \label{a1}
\end{equation}%
with a U(1) gauge field $A_{t}=\frac{J}{2r^{2}}$ and a dilaton field $\Phi
=r $, where $M$ and $J$ are the mass and charge of the (1+1)-dimensional
charged black hole. Supposing that the black hole has a positive mass and
the charge $J$ is nonzero, there are two horizons of the Achucarro-Ortiz
black hole 
\begin{equation}
r_{\pm }^{2}=\frac{M\pm \sqrt{M^{2}-\Lambda J^{2}}}{2\Lambda },  \label{a2}
\end{equation}%
where $r_{\pm }$ is the outer and inner horizon respectively. The Hawking
temperature $T_{H}$ of the event horizon is \cite{Kumar} 
\begin{eqnarray}
T_{H} &=&\frac{\sqrt{2\Lambda }}{2\pi }\frac{\sqrt{M^{2}-\Lambda J^{2}}}{(M+%
\sqrt{M^{2}-\Lambda J^{2}})^{\frac{1}{2}}}  \notag \\
&=&\frac{\Lambda }{2\pi }(\frac{r_{+}^{2}-r_{-}^{2}}{r_{+}}),  \label{a3}
\end{eqnarray}%
and the Ricci scalar is 
\begin{equation}
R=-\bigg[2\Lambda +\frac{3J^{2}}{2r^{4}}\bigg].
\end{equation}

We can rewrite the line element (\ref{a1}) in a conformal form 
\begin{equation}
ds^{2}=\Omega (r)(-dt^{2}+dr\ast ^{2}),  \label{a4}
\end{equation}%
in which 
\begin{equation}
\Omega =(-M+\Lambda r^{2}+\frac{J^{2}}{4r^{2}}),\quad \frac{dr}{dr\ast }%
=\Omega .
\end{equation}%
Then the nonzero Christoffel symbols of the metric (\ref{a4}) are 
\begin{equation}
\Gamma _{tt}^{r\ast }=\Gamma _{tr\ast }^{t}=\Gamma _{r\ast r\ast }^{r\ast }=%
\frac{1}{2}\frac{d}{dr}\Omega .  \label{a5}
\end{equation}

\section{Trace anomaly}

We consider a massless scalar field in (1+1)-dimensional Achucarro-Ortiz
black hole background. The stress tensor is covariantly conserved 
\begin{equation}
\nabla _{\nu }\Sigma _{\mu }^{\nu }=0,\qquad \Sigma _{\mu }^{\mu }=\frac{1}{%
24\pi }R,  \label{b1}
\end{equation}%
in which $\nabla $ is the covariant differentiation operator, $\Sigma _{\mu
}^{\mu }$ is the trace anomaly coming from the renormalization process, and $%
R$ is the curvature scalar.

We assume the background spacetime is steady, then the stress tensor $\Sigma
_{\mu }^{\nu }$ is time independent. Considering the nonzero Christofell
symbols (\ref{a5}), the covariantly conserved equation (\ref{b1}) in this
background takes the form 
\begin{eqnarray}
\partial _{r\ast }\Sigma _{t}^{r\ast }+\Gamma _{tr\ast }^{t}\Sigma
_{t}^{r\ast }-\Gamma _{tt}^{r\ast }\Sigma _{t}^{r\ast } &=&0,  \label{b2} \\
\partial _{r\ast }\Sigma _{r\ast }^{r\ast }+\Gamma _{tr\ast }^{t}\Sigma
_{r\ast }^{r\ast }-\Gamma _{r\ast t}^{t}\Sigma _{t}^{t} &=&0,  \label{b3}
\end{eqnarray}%
with $\Sigma _{r\ast }^{t}=-\Sigma _{t}^{r\ast },$ and $\Sigma
_{t}^{t}=\Sigma _{\alpha }^{\alpha }-\Sigma _{r\ast }^{r\ast }$. Using Eqs. (%
\ref{a5}), (\ref{b2}), (\ref{b3}) one can get that 
\begin{eqnarray}
\frac{d}{dr}(\Omega (r\ast )\Sigma _{t}^{r\ast }) &=&0,  \label{b4} \\
\frac{d}{dr}(\Omega (r\ast )\Sigma _{r\ast }^{r\ast }) &=&\frac{1}{2}\{\frac{%
d}{dr}(\Omega (r\ast )\}.  \label{b5}
\end{eqnarray}%
Eq.(\ref{b4}) leads to 
\begin{equation}
\Sigma _{t}^{r\ast }=\alpha \Omega ^{-1}(r),
\end{equation}%
where $\alpha $ is an integration constant. And the solution of Eq.(\ref{b5}%
) can be written as following 
\begin{equation}
\Sigma _{r\ast }^{r\ast }=[H(r)+\beta ]\Omega ^{-1},\qquad \beta =\Omega
(L)\Sigma _{r\ast }^{r\ast }(L),
\end{equation}%
where 
\begin{equation}
H(r)=\frac{1}{2}\int_{L}^{r}\Sigma _{\alpha }^{\alpha }{r^{\prime }}\frac{d}{%
dr^{\prime }}\Omega (r^{\prime })dr^{\prime },  \label{b6}
\end{equation}%
in which $L$ is an arbitrary scalar length. Given different scalar length $L$%
, we can get the different contribution of $H(r)$ to the stress tensor. We
need to choose $L$ more carefully. It should not include the region that is
very near the horizon because there are quantum fluctuation in this area 
\cite{Ghafarnejad}. We consider that the region out of event horizon can be
separated into two areas, one is near the horizon $r_{+}<r<(L=r_{+}+%
\varepsilon )$ where $\varepsilon $ is a very small length, and the other is 
$L<r<\infty $.

Using Equations(\ref{b4}) and (\ref{b5}), one can obtain 
\begin{equation}
\Sigma _{\mu }^{\nu }=\bigg(%
\begin{array}{cc}
\Sigma _{\alpha }^{\alpha }(r)-\Omega ^{-1}(r)H(r) & 0 \\ 
0 & \Omega ^{-1}H(r)%
\end{array}%
\bigg )+\Omega ^{-1}(r)\bigg(%
\begin{array}{cc}
-\beta & -\alpha \\ 
\alpha & \beta%
\end{array}%
\bigg ).  \label{b7}
\end{equation}%
The stress tensor $\Sigma _{\mu }^{\nu }$ can be written in this form 
\begin{equation}
\Sigma _{\mu }^{\nu }=\Sigma _{\mu }^{(1)\nu }+\Sigma _{\mu }^{(2)\nu
}+\Sigma _{\mu }^{(3)\nu },
\end{equation}%
where 
\begin{eqnarray}
\Sigma _{\mu }^{(1)\nu } &=&\bigg(%
\begin{array}{cc}
\Sigma _{\alpha }^{\alpha }(r)-\Omega ^{-1}(r)H(r) & 0 \\ 
0 & \Omega ^{-1}H(r)%
\end{array}%
\bigg ),  \label{b8} \\
\Sigma _{\mu }^{(2)\nu } &=&K\Omega ^{-1}(r)\bigg(%
\begin{array}{cc}
-1 & 0 \\ 
0 & 1%
\end{array}%
\bigg ),  \label{b9} \\
\Sigma _{\mu }^{(3)\nu } &=&\alpha \Omega ^{-1}(r)\bigg(%
\begin{array}{cc}
1 & -1 \\ 
1 & -1%
\end{array}%
\bigg ),  \label{b10}
\end{eqnarray}%
where $K=(\alpha +\beta )$. The next step is to confirm the value of $\alpha 
$ and $K$. We subject the stress tensor to the condition on which the
background spacetime is quasi-flat $r_{qf}$. We can get $\Omega
(r_{qf})\rightarrow 1,\Sigma _{\alpha }^{\alpha }(r_{qf})\rightarrow 0$ in
this region. One can compare the energy density of a beam of black body (the
Achucarro-Ortiz black hole) radiation in the quasi-flat region with the one
of equilibrium gas with Hawking temperature $T_{H}$. The stress tensor of
this equilibrium gas is \cite{Christensen} 
\begin{equation}
T_{\mu }^{(e)\nu }=\frac{\pi }{12}(kT_{H})^{2}\bigg(%
\begin{array}{cc}
-2 & 0 \\ 
0 & 2%
\end{array}%
\bigg).  \label{b11}
\end{equation}%
Comparing the stress tensor $\Sigma _{\mu }^{(2)\nu }(r_{qf})$ with Eq.(\ref%
{b11}), we can get 
\begin{equation}
K=\frac{\pi }{6}(kT_{H})^{2}.  \label{b12}
\end{equation}%
The outwards flux of thermal radiation in quasi-flat region can be described
by the stress tensor 
\begin{equation}
T_{\mu }^{(r)\nu }=\frac{\pi }{12}(kT_{H})^{2}\bigg(%
\begin{array}{cc}
-1 & -1 \\ 
1 & 1%
\end{array}%
\bigg).  \label{b13}
\end{equation}%
And the density and the flux are actually equal for a massless field, so
that 
\begin{equation}
\alpha =\frac{1}{2}[H(r_{qf})-\Sigma _{\alpha }^{\alpha }(r_{qf})].
\label{b14}
\end{equation}%
Under Eqs. (\ref{b12}) and (\ref{b14}), and considering $K=\alpha +\beta $
and $\beta =\Omega (L)\Sigma _{r^{\ast }}^{r^{\ast }}(L)$, one can obtain 
\begin{equation}
\frac{1}{2}[H(r_{qf})-\Sigma _{\alpha }^{\alpha }(r_{qf})]=\frac{\pi }{6}%
(kT_{H})^{2}-\Omega (L)\Sigma _{r^{\ast }}^{r^{\ast }}(L).  \label{b15}
\end{equation}%
We find that the total stress tensor $\Sigma _{\mu }^{\nu }$ is the function
of $\Sigma _{\alpha }^{\alpha },r_{qf}$ and $L$. $\Sigma _{\alpha }^{\alpha
}=\frac{R}{24\pi },r_{qf}$ is determined by $g_{tt}=1$, we also can get the
value of $L$ from Eq.(\ref{b15}). The stress tensor in this region can be
connected to the equilibrium gas of a background heat bath at the
temperature of $T_{H}$ and the trace anomaly.

\section{Gauge anomaly and gravitational anomaly}

In this section, we will discuss the stress tensor in the region $r\in
\lbrack r_{+},r_{+}+\varepsilon ]$. There are gauge and gravitational
anomalies in this region. We consider the current charge and gauge anomaly
at first. The current shows an anomaly in the region near the horizon if we
omit the ingoing modes in this area. The consistent form of $d$=2 Abelian
anomaly of right-handed fields is given by \cite{Bertlmann1} 
\begin{equation}
\nabla _{\mu }J^{\mu }=-\frac{e^{2}}{4\pi \sqrt{-g}}\epsilon ^{\mu \nu
}\partial _{\mu }A_{\nu },
\end{equation}%
in which 
\begin{equation}
\epsilon ^{\mu \nu }=\bigg(%
\begin{array}{cc}
0 & 1 \\ 
-1 & 0%
\end{array}%
\bigg).
\end{equation}%
But the current $J$ is not covariant, we need to define a new covariant
current \cite{Bardeen} 
\begin{equation}
\widetilde{J}^{\mu }=J^{\mu }+\frac{e^{2}}{4\pi \sqrt{-g}}A_{\lambda
}\epsilon ^{\lambda \mu }.
\end{equation}

One can find that the coefficient of $\widetilde{J}$ is twice of $J$. The
gauge is anomalistic in the region $r\in \lbrack r_{+},r_{+}+\epsilon ]$,
the current satisfies 
\begin{equation}
\partial _{r}J_{(H)}^{r}=\frac{e^{2}}{4\pi }\partial _{r}A_{t},
\end{equation}%
and it is conserved $\partial _{r}J_{(o)}^{r}=0$ outside the horizon 
\begin{equation}
\partial _{r}J_{(o)}^{r}=0.
\end{equation}%
So we can obtain 
\begin{eqnarray}
J_{(H)}^{r} &=&c_{H}+\frac{e^{2}}{4\pi }\left( A_{t}(r)-A_{t}(r_{+})\right) ,
\\
J_{(o)}^{r} &=&c_{o},
\end{eqnarray}%
where $c_{o}$ and $c_{H}$ are integration constants.

Under gauge transformations, considering the ingoing mode near the horizon,
variation of the effective action is given by $-\delta W=\int d^{2}x\sqrt{%
-g_{(2)}}\lambda \nabla _{\mu }J_{(2)}^{\mu },$ with a gauge parameter $%
\lambda $. We separate the current as two parts $J^{\mu }=J_{(o)}^{\mu
}\Theta _{+}(r)+J_{(H)}^{\mu }H(r),$ where $J_{(o)}^{\mu }$ is the current
of out horizon and $J_{(H)}^{\mu }$ is the one near horizon, and $\Theta
_{+}(r)=\Theta (r-r_{+}-\epsilon ),H(r)=1-\Theta _{+}(r)$. The variation
near the horizon becomes 
\begin{equation}
0=-\delta W=\int d^{2}x\lambda \lbrack \delta (r-(r_{+}+\epsilon
))(J_{o}^{r}-J_{H}^{r}+\frac{e^{2}}{4\pi }A_{t})+\partial _{r}(\frac{e^{2}}{%
4\pi }A_{t}H)].
\end{equation}%
The last term is canceled by quantum effects, and it is called the
Wess-Zumino term induced by the ingoing modes near the horizon \cite{Iso}.
For the delta-function is non-zero in the region $r\in \lbrack
r_{+},r_{+}+\epsilon ]$, the coefficient of this term should vanish 
\begin{equation}
c_{o}=c_{H}-\frac{e^{2}}{4\pi }A_{t}(r_{+}),
\end{equation}%
where $c_{H}$ is the value of the consistent current at the horizon.

It is necessary to fix the value of current at the horizon to determine the
current flow. Since the condition should be gauge covariant, we impose that
the coefficient of the covariant current at the horizon should vanish.
Considering $\tilde{J^{r}}=J^{r}+\frac{e^{2}}{4\pi }A_{t}(r)H(r)$, the value
of the charge flux should be 
\begin{equation}
c_{o}=-\frac{e^{2}}{2\pi }A_{t}(r_{+})=\frac{e^{2}Q}{4\pi r_{+}^{2}}.
\label{Jflux}
\end{equation}

We now turn to discuss the gravitation anomaly. Remember the $(1+1)$%
-dimensional gravitational anomaly can be described by Eq.(\ref{eq1}).
Similar as the calculation of gauge anomaly, we can get the flux of stress
tensor near the horizon. The total flux of stress tensor is 
\begin{equation}
a_o=\frac{e^2Q^2}{4\pi r_+^2} +N^r_t(r_+) =\frac{e^2Q^2}{16\pi r_+^2} +\frac{%
\pi }{12 \beta^2} .  \label{gravit}
\end{equation}

We want to find the relation between anomalies near horizon and Hawking
radiation. One can consider the flux from black body radiation (Hawking
radiation of Achucarro-Ortiz black hole) in the positive $r$ direction with
a chemical potential. The Planck distribution in Achucarro-Ortiz black hole
is given by \cite{Iso} 
\begin{subequations}
\begin{eqnarray}
I^{(\pm )}(w) &=&\frac{1}{e^{\beta (w\pm c)}-1}, \\
J^{(\pm )}(w) &=&\frac{1}{e^{\beta (w\pm c)}+1},
\end{eqnarray}%
for bosons and fermions respectively, where $c=-eA_{t}(r_{+})$ \cite%
{Hawking1}. $I^{(-)}$ and $J^{(-)}$ correspond to the distributions for
particles with charge $e$. If $(w\pm c)>0$, those distributions are
suppressed exponentially, but if $(w\pm c)<0$, we must discuss more deeply.
When $(w\pm c)<0,$ for bosons superradiance will appear, but for fermions,
detailed calculations show that there is no superradiance \cite{Unruh2}. We
will calculate the flux of fermion for simplicity. 
\end{subequations}
\begin{eqnarray}
J^{r} &=&e\int_{0}^{\infty }\frac{dw}{2\pi }(J^{-}(w)-J^{+}(w))=\frac{e^{2}Q%
}{4\pi r_{+}^{2}},  \label{flux1} \\
T_{t}^{r} &=&\int_{0}^{\infty }\frac{dw}{2\pi }w(J^{-}(w)+J^{+}(w))=\frac{%
\pi }{12\beta ^{2}}+\frac{e^{2}Q^{2}}{16r_{+}^{4}}.  \label{flux2}
\end{eqnarray}%
Compering Eq.(\ref{flux1}) with Eq.(\ref{Jflux}), and Eq.(\ref{flux2}) with
Eq.(\ref{gravit}), we find that the flux derived from blackbody radiation at
Hawking temperature cancels the flux of charge and energy.

\section{Conclusions and comments}

We have studie the contribution of different anomalies to stress tensor in
different regions in $(1+1)$-dimensional Achucarro-Ortiz black hole. In the
region of $L<r<\infty $, we find the contribution of trace anomaly to the
stress tensor can be connected to the equilibrium gas of a background heat
bath at the temperature of $T_{H}$. But in the region near the horizon,
gauge anomaly and gravitational anomaly should be considered. To avoid
losing the general covariance and gauge invariance at quantum level, the
total flow of charge and energy-momentum near the horizon of Achucarro-Ortiz
black hole must be equal to the one of blackbody radiation at Hawking
temperature. This is different from \cite{setare2} in which the author
integrates out modes in a sandwich surrounding the horizon just like \cite%
{Robinson}. However, we can get the same result (\ref{Jflux}) and (\ref%
{gravit}) as \cite{setare2} in which the BTZ black hole is reduced to
(1+1)-dimensional black hole.

Anomaly is a key to the deeper understanding of quantum field theory. In 
\cite{setare}, the relation between the trace anomaly and Casimir effect is
discussed in (1+1) dimensional curved spacetime. There are also many papers
in which the relation between black hole entropy and anomaly is discussed 
\cite{Amelino,carlip,Dabholkar,Kraus}. Nugayer \cite{Nugayer} had studied
the relation between Casimir effect and black hole evaporation two decades
ago. Many other people discussed the connection of Casimir effect and black
hole thermodynamics \cite{Hartnoll,Belgiorno,Widom,Sassaroli,Yun}. The
relation between them will be studied more deeply and be understood more
clearly.

\begin{acknowledgments}
Xiao thanks Vagenas, Robinson, Iso, Umetsu and Setare for the useful
discussions about their works. We would also like to thank the referee of
this paper for pointing out many errors in the original manuscript.

K. Xiao and W. B. Liu are supported by NBRPC ( Grant No.2003CB716302 ) ,
NSFC ( Grant No.10475013 ) and H. B. Zhang is supported by NSFC ( Grant
No.10533010 ) .
\end{acknowledgments}

\end{document}